# Defect scattering in graphene

Jian-Hao Chen[1,2], W. G. Cullen[2], C. Jang[2], and M. S. Fuhrer[1,2] and E. D. Williams[1,2]

[1]*Materials Research Science and Engineering Center and* [2]*Center for Nanophysics and Advanced Materials, Department of Physics, University of Maryland, College Park, MD 20742 USA*

Abstract

Irradiation of graphene on $SiO_2$ by 500 eV Ne and He ions creates defects that cause intervalley scattering as evident from a significant Raman *D* band intensity. The defect scattering gives a conductivity proportional to charge carrier density, with mobility decreasing as the inverse of the ion dose. The mobility decrease is four times larger than for a similar concentration of singly charged impurities. The minimum conductivity decreases proportional to the mobility to values lower than $4e^2/\pi h$, the minimum theoretical value for graphene free of intervalley scattering. Defected graphene shows a diverging resistivity at low temperature, indicating insulating behavior. The results are best explained by ion-induced formation of lattice defects that result in mid-gap states.

The strong carbon-carbon $sp^2$ bonds which provide graphene with high intrinsic strength [1] and make possible the isolation of single atomic layers [2], also result in a very low density of lattice defects in graphene prepared by mechanical exfoliation [3, 4]. However, lattice defects in graphene are of great theoretical interest [5, 6] as a potential source of intervalley scattering, which in principle transforms graphene from a metal to an insulator [7, 8]. Lattice defects are also likely to be present in various concentrations in graphene synthesized by reduction of graphene oxide [9, 10], chemical vapor deposition [11, 12], or segregation of carbon on the surface of SiC [13], hence it is important to understand their impact on electronic transport.

Here we show that ion irradiation-induced defects in graphene cause a significant intensity in the Raman *D* band associated with intervalley electron scattering [14-16] and give rise to a constant mobility, similar to the effect of charged impurities, but with a magnitude 4 times lower than for a similar concentration of singly charged impurities. This result is in contrast to the carrier-density-independent conductivity for weak point disorder[17, 18] but consistent with the theory of strong scattering by mid-gap states[5, 6]. Unlike charged impurities [19], lattice defects (1) do not change the residual charge density in electron-hole puddles; (2) greatly depress the minimum conductivity, even below $4e^2/\pi h$ (the theoretical minimum value of the conductivity at the Dirac point in the absence of intervalley scatteing [7]); and (3) induce insulating temperature dependence of the conductivity.

Transport with constant mobility is predicted for both charged impurity scattering and scattering by mid-gap states. Charged-impurity disorder in graphene results in a conductivity

$$\sigma_c = ne\mu_c = \frac{2e^2}{h}\frac{n}{n_c}\frac{1}{G(2r_s)} \qquad (1)$$

where $e$ is the electronic charge, $h$ the Planck's constant, $n_c$ the charged impurity density, $r_s$ the Wigner-Seitz radius and $G(2r_s)$ an analytical function of the dimensionless interaction strength in graphene. For graphene on $SiO_2$, Eq. (1) gives $\mu_c \approx 5 \times 10^{15}$ $V^{-1}s^{-1}/n_c$[19, 20]. The random charged impurity potential also gives rise to electron-hole puddles with a characteristic intrinsic carrier density $n^*$, which is a function only of $n_c$, $d$ (the impurity-graphene distance) and $r_s$, resulting in a minimum conductivity $\sigma_{min} = n^*e\mu_c$. However, strong disorder, modeled as a deep potential well of radius $R$, is predicted to produce midgap states in graphene[6], and a conductivity which is also roughly linear in $n$[5]:

$$\sigma_d = ne\mu_d = \frac{2e^2}{\pi h}\frac{n}{n_d}\ln^2(k_F R) \qquad (2)$$

where $n_d$ is the defect density and $k_F$ is the Fermi wavevector. A third type of scattering in graphene, weak point disorder, is predicted to give rise to a carrier-density-independent resistivity $\rho_s$[17], which has been observed experimentally[18].

To investigate the dependence of graphene's conductivity on defect density, cleaned graphene on $SiO_2$ was irradiated with 500 eV $He^+$ and $Ne^+$ ions in ultra-high vacuum (UHV) at low temperature (10K for $He^+$ irradiation and 40-80K for $Ne^+$ irradiation)[21]. Ion irradiation of graphite at these energies produces one atomic-scale defect, most likely a carbon vacancy with a trapped rare-gas molecule, per incident ion [22, 23].

Figure 1 shows the Raman spectrum, taken under ambient conditions, for a representative graphene sample before irradiation, and after irradiation by $Ne^+$ at a dose of $10^{12}$ $cm^{-2}$ (~ 1 $Ne^+$ per $4 \times 10^3$ carbon atoms). The pristine sample shows a Lorentzian $G'$ band characteristic of single layer graphene, and no detectable $D$ band. Upon irradiation, the appearance of the $D$ band indicates significant intervalley scattering [14, 16]. A very rough estimate of the defect spacing can be made using the empirical formula $L_a = \left[2.4 \times 10^{-10} \text{nm}^{-3}\right] \lambda^4 \left(\frac{I_D}{I_G}\right)^{-1}$, which relates the grain size $L_a$ in disordered graphite, to the ratio of the integrated D and G band intensities $I_D$ and $I_G$, and $\lambda$ the excitation wavelength (633 nm) [24]. Applying this formula to our irradiated graphene gives $L_a \sim 60$ nm, larger than the expected defect spacing of 10 nm, but comparable to the transport mean free path of ~50 nm (see below).

Figure 2 shows the $\sigma(V_g)$ curves for the pristine sample and 5 different Ne$^+$ irradiation concentration at $T = 41$K in UHV as well as predictions from Eq.(2) with the experimentally extracted defect radius $R$ at $n_d = 7.22 \times 10^{11}$cm$^{-2}$ (see below). $\mu$ and $\sigma_{min}$ partially recover after heating to 485K between runs, possibly due to annealing or passivation of the defects. To determine $\mu$, and the resistivity $\rho_s$ due to weak point disorder, the $\sigma(V_g)$ curves are fitted to the form $\sigma(V_g)^{-1} = \left[c_g\left(V_g - V_{g,min}\right)\mu\right]^{-1} + \rho_s$ [18]. We fit the hole side of the $\sigma(V_g)$ curve ($V_g < V_{g,min}$) because the data span a wider $V_g$ range. Figure 3a shows $1/\mu$ vs. ion dosage for four experimental runs on two different graphene samples as well as behavior for charged impurities[19, 20]. For the irradiated samples, $1/\mu$ increases linearly with ion dosage as expected for uncorrelated scattering. Fitting yields a proportionality of $7.9 \times 10^{-16}$ Vs for the Ne$^+$ irradiation runs and $9.3 \times 10^{-16}$ Vs for the He$^+$ irradiation runs and an offset that yields the intrinsic defect density of the graphene prior to irradiation[21]. Assuming mid-gap scattering (Eq. 2), at carrier density $n = 2 \times 10^{12}$ cm$^{-2}$, the proportionality constant yields the defect radius $R = 2.3$ Å for Ne$^+$ irradiation and 2.9 Å for He$^+$ irradiation. If the proportionality is attributed to charged defect scattering (Eq. 1), it would require addition of charge $Z \sim 4e$ per incident ion. Figure 3b shows the density-independent resistivity $\rho_s$ for the same four experimental runs; $\rho_s$ is very small (on order $10^{-3}$ h/e$^2$) and does not change significantly with ion irradiation dose.

Figure 4a shows the change in the voltage of the minimum conductivity $\Delta V_{g,min}$ as a function of the inverse mobility $1/\mu$ (proportional to ion dose) for the four ion irradiation runs. For comparison, the magnitude of $\Delta V_{g,min}$ for a potassium (K) dosing (addition of charged impurities) run is also shown (data from Ref.[19]), which is 5 times larger than a similar concentration of ion irradiation. Note that $\Delta V_{g,min}$ is positive for ion irradiation, and negative for K dosing. Figure 4b shows $\sigma_{min}$ vs. $\mu$ for the same four ion irradiation runs and the K dosing run [19]. In sharp contrast to the charged impurities introduced by K dosing, where $\sigma_{min} = n^*e\mu_c$ varies slowly and non-monotonically because $n^*$ increases with increasing dose (decreasing $\mu$), ion irradiation has a large effect on $\sigma_{min}$, reducing $\sigma_{min}$ roughly proportional to $\mu$.

We now discuss the changes in $\sigma(n)$ upon ion irradiation. The density-independent resistivity (Fig. 3b) $\rho_s \sim 3\times 10^{-3}$ $h/e^2$ and is roughly independent of ion dose; at a carrier density of $10^{12}$ cm$^{-2}$, this corresponds to a mean free path >2 µm. The dominant signature, linear $\sigma(n) = ne\mu_d$ with $\mu_d$ independent of $n$, indicates that ion irradiation either creates mid-gap states or charged impurities. However, several observations argue that the observed changes in $\sigma(n)$ are dominated by lattice defects: (1) The intervalley scattering observed in Raman spectroscopy (Fig. 1) with scattering length on order 60 nm is inconsistent with $\rho_s$, but consistent with the associated mobility $\mu = 1300$ cm$^2$ V$^{-1}$ s$^{-1}$, at an ambient doping level of $\sim 10^{13}$ cm$^{-2}$, from which we calculate a mean free path $l \sim 50$ nm. This correspondence suggests that the transport mean free path significantly probes intervalley scattering from lattice defects. (2) The sign $\Delta V_{g,min}$ for ion irradiation is positive, opposite to the expectation for deposition of positive ions near the graphene and also opposite to what was observed for ion-irradiated MOSFETs [25]. (3) The reduction in mobility, if due to charged impurities, would require ~4 added charges per incident ion, while $\Delta V_{g,min}$ indicates only ~1/5 of a net charge per incident ion; this would require a delicate balance between creation of positive and negative impurities, and such balance would need to hold for incident Ne$^+$ and He$^+$, which have very different momenta. (4) Within the

Boltzmann transport picture, $\sigma_{min} = n^*e\mu$ [20] where the total mobility $\mu = (\mu_d^{-1} + \mu_c^{-1})^{-1}$. The roughly proportional relationship between $\sigma_{min}$ and $\mu$ for ion-irradiated samples indicates that $n^*$, which is a function of $n_c$, is nearly independent of ion dose [26].

We therefore conclude that the data of Figure 3a are dominated by uncharged lattice defects in graphene. The impurity radius $R \sim 2.3$ Å – 2.9 Å obtained from the linear fits of Fig. 3 is a reasonable value for single-carbon vacancies generated by ion knock-off [23]. Using this value of $R$ in Eq. (2) yields a $\sigma(V_g)$ similar in magnitude to the experimental curve, but with a stronger sublinearty (Fig. 2). We do not understand this discrepancy, but it may be related to carrier density inhomogeneity persisting to carrier densities much larger than $n^*$ [27], or to the addition of a small amount of deep charged impurities [25] which would contribute a supralinear $\sigma(V_g)$.

Lastly we discuss the possibility of a metal-insulator transition in graphene with defects. Disorder-free graphene is expected to have a minimum conductivity of $4e^2/\pi h$ [7]. The introduction of intravalley scattering only (e.g. charged impurities) is expected to induce weak *anti*-localization, *increasing* the conductivity[7, 8] with decreasing temperature. However, intervalley scattering (which gives rise to the Raman *D* band) is expected to induce weak localization, and insulating behavior, i.e. $\sigma \rightarrow 0$ as $T \rightarrow 0$, in graphene[7, 8]. From Fig. 4a, we can see that $\sigma_{min}$ in ion-irradiated samples can be reduced well below $4e^2/\pi h$, the minimum metallic value. Figure 5 shows the conductivity of the Ne$^+$ irradiated graphene sample as a function of temperature for three different gate voltages. The *T*-dependent conductivity of pristine graphene from Ref. [28] is also shown for comparison. The pristine graphene has metallic behavior, e.g., $d\sigma/dT < 0$. However, even a small amount of irradiation (that changes the room-temperature mobility < 4×) drastically affects the low-temperature behavior. In stark contrast to graphene without irradiation, where $\sigma_{min}$ is largely temperature independent from $T = 4\text{-}100$ K [29], our irradiated sample is insulating with diverging resistivity as $T \rightarrow 0$. More work is needed to understand the exact nature of the insulating state in ion-irradiated graphene, but the data are consistent with the expectation that intervalley scattering produces localization[7].

In conclusion, we have measured charge transport in graphene with defects induced by ion irradiation in ultra high vacuum. Defects cause significant intervalley scattering, as seen in a prominent Raman *D* band. Defects give rise to a constant mobility, with a magnitude ~4× lower than for similar concentration of potassium ions on graphene, and consistent with scattering by midgap states. In contrast to charge impurity disorder, lattice defects reduce the minimum conductivity dramatically, and produce an insulating temperature dependence of the conductivity.

Acknowledgements

This work has been supported by the NSF-UMD-MRSEC grant DMR 05-20471 (J.H.C., W.G.C, E.D.W., M.S.F.) and the US ONR grant N000140610882 (C.J., M.S.F.). The MRSEC SEFs were used in this work. Infrastructure support has also been provided by the UMD NanoCenter and CNAM.

Figure Captions

Figure 1. Raman spectra (wavelength 633 nm) for (a) pristine graphene and (b) graphene irradiated by 500 eV $Ne^+$ ions at a dose of $10^{12}$ $cm^{-2}$.

Figure 2. Conductivity vs. gate voltage curves for pristine graphene and 5 different $Ne^+$ ion irradiation concentrations at $T = 41K$ in ultra high vacuum, as well as predictions from Eq.(2) with the experimentally extracted defect radius $R=2.3$Å at defect density $n_d = 7.22\times10^{11} cm^{-2}$.

Figure 3. (a) Inverse of mobility ($1/\mu$) vs. ion dosage for two $Ne^+$ irradiation runs on sample 1 and two $He^+$ irradiation runs on sample 2. Dashed line is behavior for the same concentration of charged impurities (potassium on graphene from Ref. [19]). (b) Density-independent resistivity $\rho_s$ vs. ion dosage.

Figure 4. (a) Magnitude of the shift in the gate voltage of minimum conductivity ($|\Delta V_{g,min}|$) vs. inverse mobility ($1/\mu$). The shift is with respect to the initial value of $V_{g,min}$, 8.8V and 6.4 V for the $Ne^+$ and $He^+$ irradiated samples respectively. (b) Minimum conductivity ($\sigma_{min}$) vs. $\mu$ for two $Ne^+$ irradiation runs on sample 1 and two $He^+$ irradiation runs on sample 2. Data for potassium dosing (Ref.[19]) are shown for comparison. $V_{g,min}$ is positive for ion irradiation, negative for K dosing.

Figure 5. Temperature dependence of the conductivity $\sigma(T)$ of pristine (open symbols) and irradiated (solid symbols) graphene at three different gate voltages. $\sigma(T)$ taken on cooling is shown for Sample 1 after Run 1 (irradiation by $Ne^+$, dose $7\times10^{11}$ $cm^{-2}$) and annealing to $T = 300$ K. $\sigma(T)$ for the pristine sample is from Ref. [28].

Figures
Figure 1

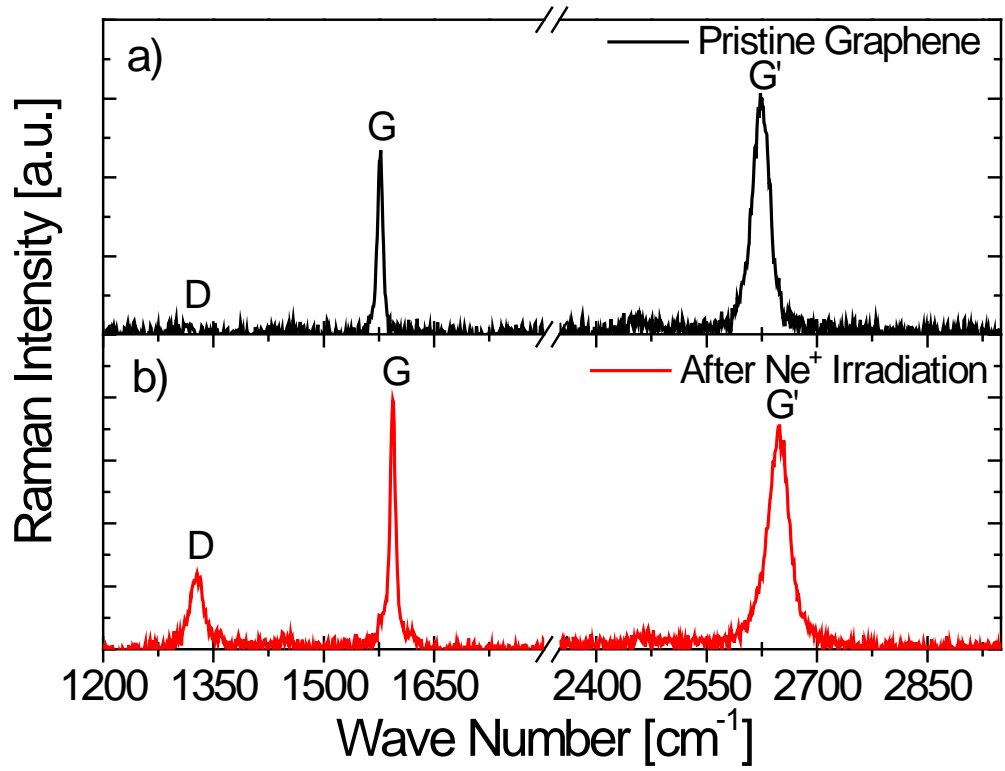

Figure 2

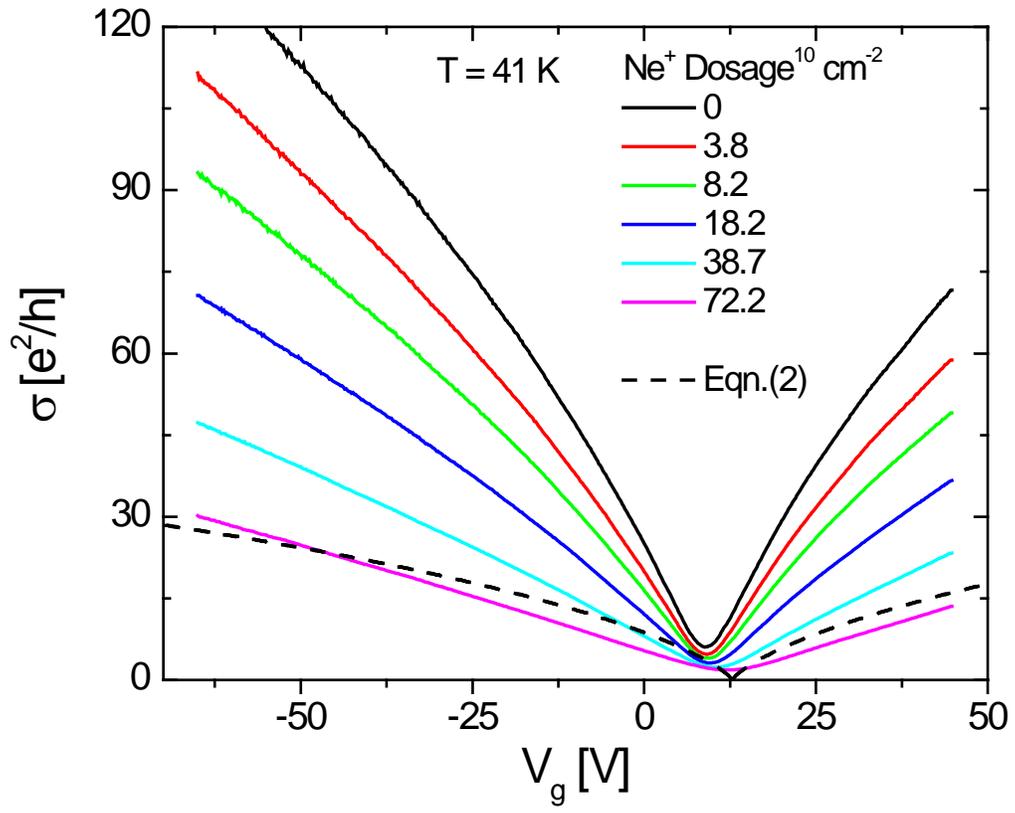

Figure 3

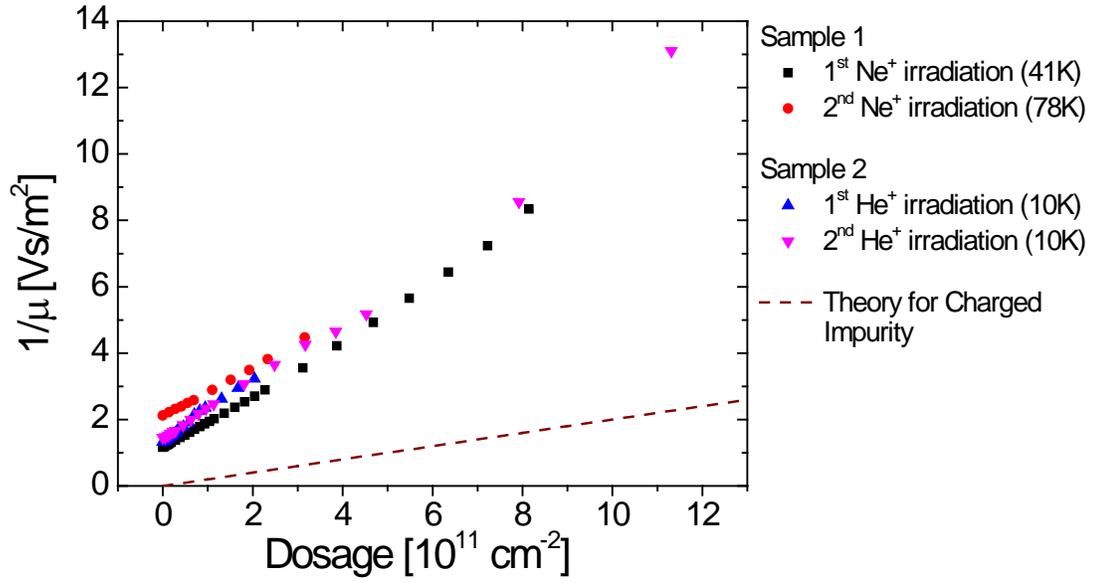

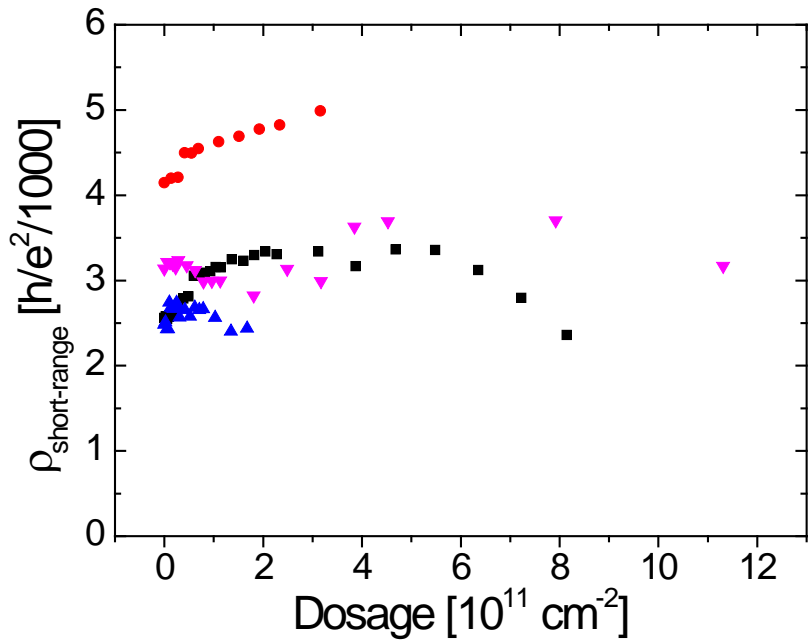

Figure 4

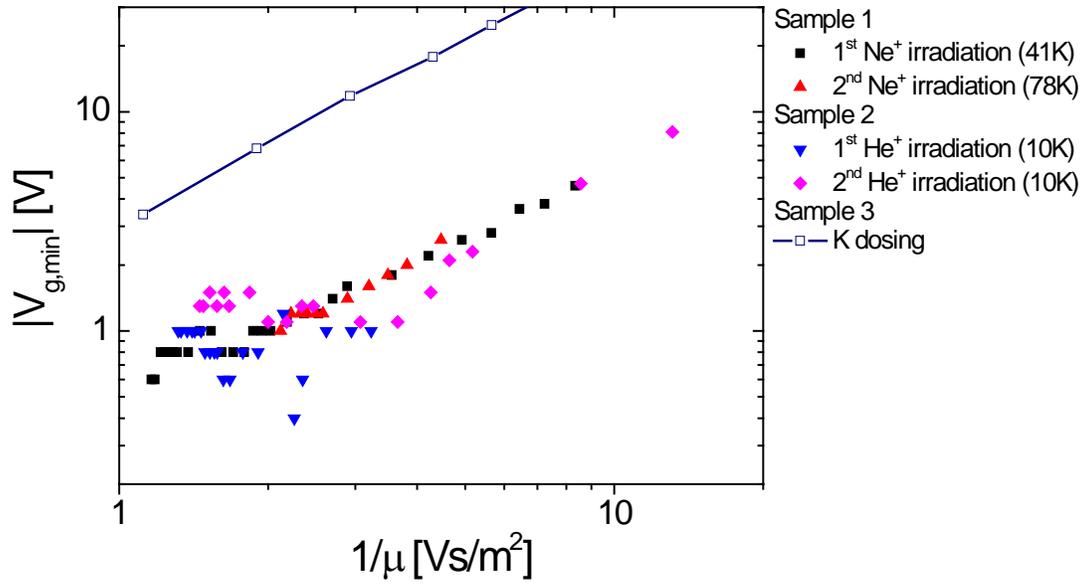

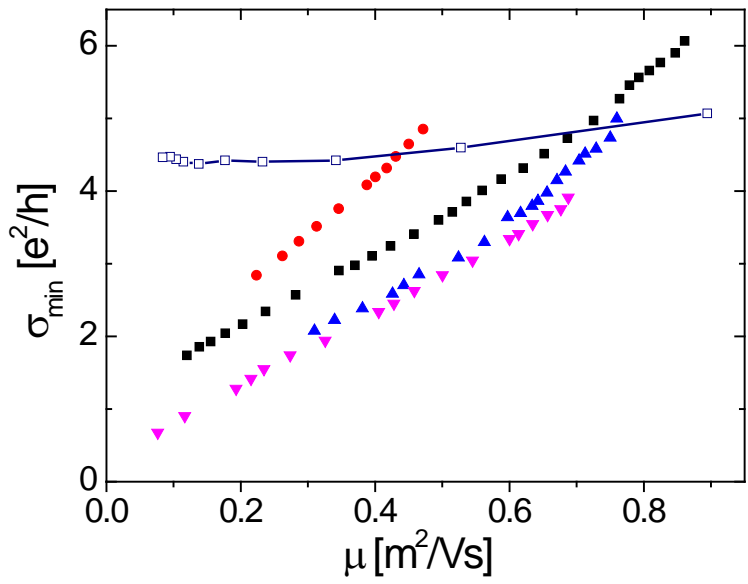

Figure 5

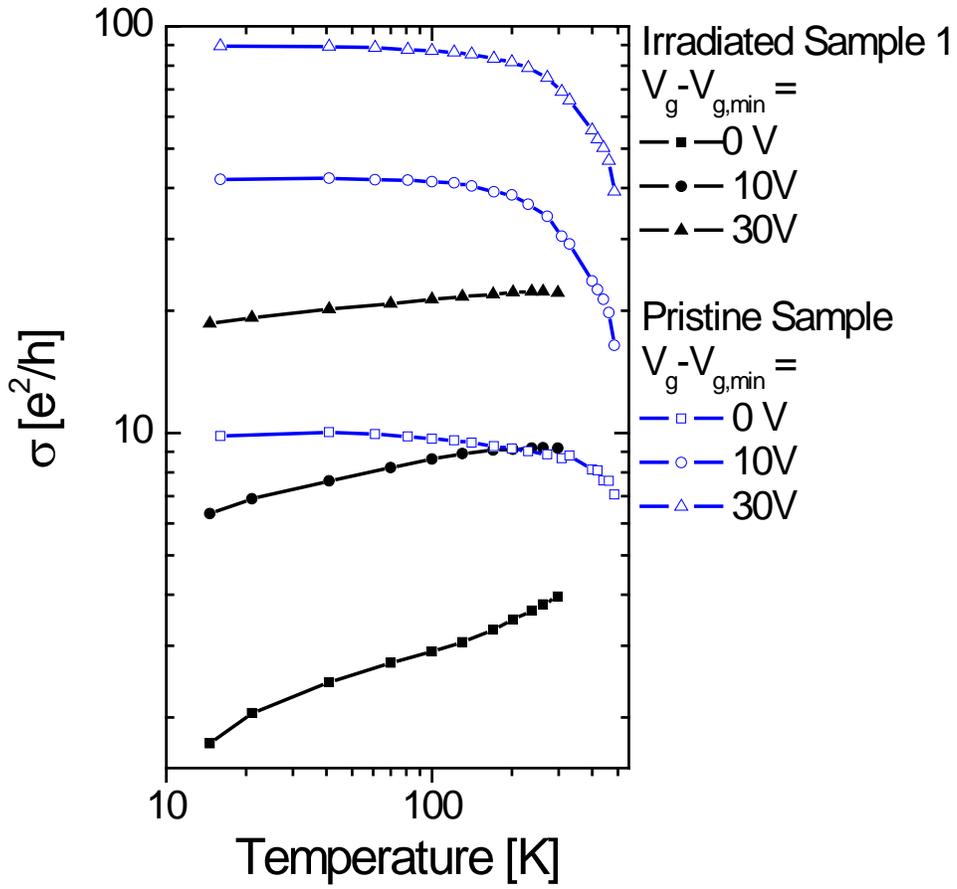

# Auxiliary Material for "Defect scattering in graphene"

Jian-Hao Chen[1,2], W. G. Cullen[2], C. Jang[2], and M. S. Fuhrer[1,2] and E. D. Williams[1,2]

[1]Materials Research Science and Engineering Center and [2]Center for Nanophysics and Advanced Materials, Department of Physics, University of Maryland, College Park, MD 20742 USA

I.      Experimental Methods

Experiments were performed on graphene obtained from Kish graphite by mechanical exfoliation [1] on 290 nm $SiO_2$ over doped Si (back gate), with Au/Cr electrodes. After fabrication, devices were cleaned in $H_2$/Ar at 300°C for 1 hour [2, 3]. The conductivity was measured in a four-probe configuration as a function of gate voltage $V_g$, where the carrier density $n = c_g V_g/e$ with $c_g = 11.9$ nF/$cm^{-2}$. Following a vacuum bakeout and overnight anneal in UHV at 490K, experiments were carried out at base pressures lower than $5\times10^{-10}$ torr and $T = 10$ K for $He^+$ irradiation and 40-80 K for $Ne^+$ irradiation, to avoid Ne adsorption on graphene. A sputter gun ionized He or Ne gas and accelerated the ions to 500 eV. A shutter controlled the irradiation time and allowed measurement of $\sigma(V_g)$ between irradiation doses. The pressure of the inert gas, up to $5*10^{-8}$ torr for Ne and up to $2.5*10^{-7}$ torr for He, was monitored by a residual gas analyzer and the ion flux calibrated by a Faraday cup mounted at the same location as the sample in a control experiment. After irradiation, each device was annealed at 485K overnight before further experimental runs were performed.

II.  Derivation of Equation 2 in main text

Eq. (2) is obtained by putting $\tau_d = \dfrac{k_F}{\pi^2 v_F n_d} \ln^2 k_F R$ (Eq. (54) from Ref. [4]) and $k_F = \sqrt{\pi n}$ into

$\sigma_d = \dfrac{2e^2}{h} k_F v_F \tau_d$, where $k_F$ is the Fermi wavevector, $v_F$ the Fermi velocity, $R$ the defect radius, $n$ the carrier density, $n_d$ the defect density and $\tau_d$ the defect scattering time.

III.  Interpretation of the offset in the linear dependence of 1/mobility on ion dosage (Fig. 3a)

Assuming the initial disorders are charged impurities, the offset yields [5] values of $n_c$ ~$4*10^{11}$ cm$^{-2}$ and ~$5*10^{11}$ cm$^{-2}$ respectively for the samples exposed to Ne$^+$ and He$^+$ ion irradiation. If such offset were ascribed to lattice defect scattering, extrapolating to $1/\mu \rightarrow 0$, it would indicate a defect concentration on order of $10^{11}$ cm$^{-2}$. However, lattice defects at this concentration should produce a prominent Raman D band, and depress the minimum conductivity. Additionally, experiments to tune the dielectric constant in graphene [6] indicate that the native impurities in exfoliated graphene are charged impurities. The mobility of 200,000 cm$^2$/Vs achieved in suspended graphene samples [7] can be used to estimate an upper bound on the native lattice defect density of exfoliated graphene of ~ $6 \times 10^9$ cm$^{-2}$.